\documentclass{amsart}
\usepackage{amsfonts}
\usepackage{amsmath,amsthm}
\usepackage{amsfonts,amssymb}
\usepackage{mathrsfs}
\usepackage{enumerate}
\usepackage{bm}

\newtheorem{thm}{Theorem}[section]

\newtheorem{lem}[thm]{Lemma}

\theoremstyle{remark}

\numberwithin{equation}{section}

\def\az{\alpha}

\def\dz{\delta}

\def\vz{\varepsilon}
\def\lz{\lambda}

\def\k{{\bm k}}

\def\va{\varepsilon}
\def\ga{\gamma}
\def\g{{\bm \gamma}}
\def\a{{\bm \alpha}}

\def\x{{\bm x}}

\def\y{{\bm y}}

\def\N{\mathbb N}

\begin{document}

\title[] {Worst case tractability of $L_2$-approximation for weighted Korobov spaces }

\author{Huichao Yan$^1$} \address{$^1$ School of Computer and Network Engineering, Shanxi Datong University, Datong 037009, China}
\email{yanhuichao@sxdtdx.edu.cn}

\author{Jia Chen$^{2,*}$} \address{$^2$ School of Mathematics and Statistics, Shanxi Datong
University, Datong 037009, China}
\email{jiachen@sxdtdx.edu.cn}



\thanks{
 $^*$ Corresponding author.}

\begin{abstract} We study $L_2$-approximation problems $\text{APP}_d$ in the worst case setting in the  weighted Korobov spaces $H_{d,\a,{\bm \ga}}$ with  parameter sequences ${\bm \ga}=\{\ga_j\}$ and $\a=\{\az_j\}$ of positive real
numbers  $1\ge \ga_1\ge \ga_2\ge \cdots\ge 0$ and $\frac1 2<\az_1\le \az_2\le \cdots$.
We consider the minimal worst case error $e(n,\text{APP}_d)$ of algorithms that use $n$   arbitrary continuous linear functionals with  $d$ variables. 
We study polynomial convergence  of the minimal worst case error, which means that $e(n,\text{APP}_d)$ converges to zero
polynomially fast with increasing $n$. We recall the notions of polynomial, strongly polynomial, weak and $(t_1,t_2)$-weak 
tractability. In particular, polynomial tractability means that we
need a polynomial number of arbitrary continuous linear functionals in $d$ and $\va^{-1}$
  with the accuracy $\va$ of the approximation. We obtain that the matching necessary
and sufficient condition on the sequences  ${\bm \ga}$ and $\a$ for strongly polynomial tractability or polynomial tractability is
$$\dz:=\liminf_{j\to\infty}\frac{\ln \ga_j^{-1}}{\ln j}>0,$$
and the exponent of strongly polynomial tractability is $$p^{\text{str}}=2\max\big\{\frac 1 \dz, \frac 1 {2\az_1}\big\}.$$

\end{abstract}

\maketitle
\input amssym.def

\section{Introduction}\label{sec1}

We investigate  multivariate approximation problems $S_d$ with large or even huge $d$. Examples include these problems in statistics, computational finance and physics. In order to solve these problems we usually consider algorithms using finitely many evaluations of arbitrary continuous linear functionals. We use either the absolute error criterion (ABS) or the normalized error criterion (NOR). For $\text{X}\in\{\text{ABS}, \text{NOR}\}$ we define the information complexity $n^\text{X}(\va,S_d)$  to be the minimal number of linear functionals  which are needed to find an algorithm whose worst case error is at most $\va$. The behavior of the information complexity $n^\text{X}(\va,S_d)$
 is the major concern when the accuracy $\va$ of approximation goes to zero and the
number $d$ of variables goes to infinity. For small $\va$ and large $d$, tractability is aimed at studying  how the information complexity $n^\text{X}(\va,S_d)$ behaves as a function of  $d$ and $\va^{-1}$, while the exponential convergence-tractability (EC-tractability)
is aimed at studying  how the information complexity $n^\text{X}(\va,S_d)$ behaves as a function of $d$ and $(1+\ln(\va^{-1}))$. Recently  the study of tractability and  EC-tractability in the worst case setting has attracted much interest in analytic Korobov spaces and weighted Korobov spaces; see \cite{DKPW, EP, IKPW, LX, NSW, NW1, NW2, NW3, W, WW, X1} and the references therein.

This paper is devoted to discussing worst case tractability  of
$L_2$-approximation problem $$\text{APP}=\big\{\textrm{APP}_d:H_{d,\a,{\bm \gamma}}\to L_2([0,1]^d)\big\}_{d\in\N}$$ with $\textrm{APP}_d(f)=f$ for all $f\in H_{d,\a,{\bm \gamma}}$ in  weighted Korobov spaces $H_{d,\a,{\bm \ga}}$ with positive parameter sequences $ \g=\{\ga_j\}_{j\in \N}$ and $\a=\{\az_j\}_{j\in \N}$. The tractability and  EC-tractability  of such problem APP with  parameters $1\ge \ga_1\ge \ga_2\ge \cdots\ge 0$ and $\frac1 2<\az_1=\az_2=\cdots$ were discussed in \cite{EP, NSW, WW} and in \cite{C}, respectively. Additionally,  \cite{EP, WW} considered the strongly polynomial tractability (SPT), polynomial tractability (PT), weak tractability (WT) and $(t_1,t_2)$-weak tractability ($(t_1,t_2)$-WT) with $t_1>1$ and $t_2>0$.

In this paper we study SPT, PT, and $(t_1,t_2)$-WT for all $t_1>1$ and $t_2>0$ of the above problem APP with  parameters $$1\ge \ga_1\ge \ga_2\ge \cdots\ge 0,$$ and $$\frac1 2<\az_1\le\az_2\le\cdots$$ for the ABS or the NOR. We obtain that the problem APP suffers from the curse of dimensionality with $\lim_{j\to\infty}\ga_j=1$. Especially, we get a sufficient and necessary condition for SPT or PT and the exponent of SPT.

The paper is organized as follows. In Section \ref{sec2.1} we give  preliminaries about multivariate  approximation problems, tractability and EC-tractability in Hilbert spaces  in the worst case setting. In Section \ref{sec2.2} we give the definitions of  weighted Korobov spaces with parameter sequences and  present the main results. Section \ref{sec3} is devoted to proving  Theorem \ref{th1}.

\

\section{Preliminaries and main results}\label{sec2}
\

\subsection{Worst case approximation and tractability in Hilbert spaces}\label{sec2.1}

\

Let $F_d$ and $G_d$ be two sequences of Hilbert spaces. Consider a sequence of compact linear operators
$$S_d:F_d\to G_d$$
for all $d\in \Bbb N$. We approximation $S_d$ by algorithm $A_{n,d}$ of the form
\begin{equation}\label{eq2.1}A_{n,d}(f)=\sum_{i=1}^n T_i(f)g_i,\,\text{for}\,\,f\in S_d,\end{equation}
where functions $g_i\in G_d$ and continuous linear functionals $T_i\in F^*_d$ for $i =1,\cdots,n$.
The worst case error for the algorithm $A_{n,d}$ of the form \eqref{eq2.1} is defined as
$$e(A_{n,d}):=\sup_{f\in F_d,||f||_{F_d}\le 1}||S_d(f)-A_{n,d}(f)||_{G_d}.$$
The $n$th minimal worst-case error, for $n\ge 1$, is defined by$$e(n,S_d):=\inf_{A_{n,d}}e(A_{n,d}),$$ where the infimum is taken over all linear algorithms of the form \eqref{eq2.1}. For $n=0$, we use $A_{0,d}=0$. We call $$e(0,S_d)=\sup_{f\in F_d,||f||_{F_d}\le1}||S_d(f)||_{G_d}$$ the initial error of the problem $S_d$.

The information complexity for $S_d$ can be studied using either the absolute error criterion (ABS), or the normalized error criterion (NOR). The information complexity $n^\text{X}(\va,S_d)$ for $\text{X}\in\{\text{ABS}, \text{NOR}\}$ is defined by
$$n^\text{X}(\va,S_d):=\min\{n\in \mathbb{N}_0:e(n,S_d)\le \va \text{CRI}_d\},$$
where \begin{equation*}
\text{CRI}_d:=\left\{\begin{matrix}
 &1, \;\; \;\;\;\;\;\;\; \text{ for}\; \text{X}=\text{ABS}, \\
 &e(0,S_d),\,\,\text{for} \ \text{X}=\text{NOR}.
\end{matrix}\right.
\end{equation*}
Here, $\mathbb{N}_0=\{0, 1, \cdots\}$ and $\Bbb N=\{1, 2, \cdots\}$.

It is well known, see e.g., \cite{NW1,TWW}, that the $n$th minimal worst case errors $e(n, S_d)$ and the information complexity $n^\text{X}(\va,S_d)$ depend on the eigenvalues of the continuously linear operator $W_d=S^*_d S_d: F_d\to F_d$. Let $(\lz_{d,j}, \eta_{d,j})$ be the eigenpairs of $W_d$, i.e.,
$$W_d\eta_{d,j}=\lz_{d,j}\eta_{d,j}\,\, \text{for all}\,\, j\in \mathbb{N},$$ where the eigenvalues $\lz_{d,j}$ are ordered,
$$\lz_{d,1}\ge \lz_{d,2}\ge \cdots \ge 0,$$
and the eigenvectors $\eta_{d,j}$ are orthonormal,
$$\langle\eta_{d,i},\eta_{d,j}\rangle_{F_d}=\delta_{i,j}\,\, \text{for all}\,\, i,\, j\in \mathbb{N}.$$
Then the $n$th minimal error is obtained for the algorithm
\begin{equation*}A_{n,d}^\diamond f=\sum_{j=1}^n\langle f, \eta_{d,j} \rangle_{F_d}\eta_{d,j}\,\, \text{for all}\,\, f\in F_d,
\end{equation*}
and
\begin{equation*}e(n, S_d)=e(A^\diamond_{n,d})=\sqrt{\lz_{d,n+1}}\,\, \text{for all}\,\, n\in \mathbb{N}_0.\end{equation*}
Hence the information complexity is equal to
\begin{align}\label{eq2.2}n^\text{X}(\va,S_d)&=\min\{n\in \mathbb{N}_0 : \sqrt{\lz_{d,n+1}}
\le \va\text{CRI}_d\}\notag\\
&=\min\{n\in \mathbb{N}_0 : \lz_{d,n+1}
\le \va^2\text{CRI}_d^2\}\notag\\
&=|\{n\in \mathbb{N}: \lz_{d,n}> \va^{2}\text{CRI}_d^2\}|,\end{align}
with $\va\in(0,1)$ and $d\in\N$.

Various notions of tractability and exponential convergence-tractability have been studied  for multivariate problems. We briefly recall some of the basic tractability and exponential convergence-tractability (EC-tractability) notions.

Let ${S}=\{S_d\}_{d\in \Bbb N}$. For $\text{X}\in\{\text{ABS}, \text{NOR}\}$, we say ${S}$ is

$\bullet$ Strongly polynomially tractable (SPT) iff there exist non-negative numbers $C$ and $p$ such that for all $d\in \mathbb N$, $\va\in(0,1)$,
$$n^\text{X}(\va,S_d)\le C(\va^{-1})^p.$$
The exponent $p^{\text{str}}$ of SPT is defined to be the infimum of all $p$ for which the above inequality holds.

$\bullet$ Polynomially tractable (PT) iff there exist non-negative numbers $C$, $p$ and $q$ such that for all $d\in \mathbb N$, $\va\in(0,1)$,
$$n^\text{X}(\va,S_d)\le Cd^q(\va^{-1})^p.$$

$\bullet$  Weakly tractable (WT) iff
$$\lim_{\va ^{-1}+d\rightarrow \infty }\frac{\ln n^\text{X}(\va ,S_d)}{d+\va
^{-1}}=0.$$

$\bullet$ $(t_1,t_2)$-weakly tractable ($(t_1,t_2)$-WT) for fixed positive $t_1$ and $t_2$ iff
$$\lim_{\varepsilon ^{-1}+d\rightarrow \infty }\frac{\ln n^\text{X}(\va,S_d)}{d^{t_1 }+(\va ^{-1})^{t_2}}=0.$$

$\bullet$ ${S}$ suffers from the curse of dimensionality if there exist positive
numbers $C_1$, $C_2$, $\va_0$ such that for all $0<\va\le \va_0$ and infinitely many $d\in \mathbb N$,
$$n(\va, d)\ge C_1(1+C_2)^d.$$

$\bullet$ Exponential convergence-strongly polynomially tractable (EC-SPT) iff there exist non-negative numbers $C$ and $p$ such that for all $d\in \mathbb N$, $\va\in(0,1)$,
$$n^\text{X}(\va,S_d)\le C\big(1+\ln(\va^{-1})\big)^p.$$
The exponent of EC-SPT is defined to be the infimum of all $p$ for which the above inequality holds.

$\bullet$ Exponential convergence-polynomially tractable (EC-PT) iff there exist non-negative numbers $C$, $p$ and $q$ such that for all $d\in \mathbb N$, $\va\in(0,1)$,
$$n^\text{X}(\va,S_d)\le Cd^q\big(1+\ln(\va^{-1})\big)^p.$$

$\bullet$ Exponential convergence-weakly tractable (EC-WT) iff
$$\lim_{\varepsilon ^{-1}+d\rightarrow \infty }\frac{\ln n^\text{X}(\va,\text{APP}_d)}{d+\ln(\va ^{-1})}=0.$$

$\bullet$ Exponential convergence-$(t_1,t_2)$-weakly tractable (EC-$(t_1,t_2)$-WT) for fixed positive $t_1$ and $t_2$ iff $$\lim_{\varepsilon ^{-1}+d\rightarrow \infty }\frac{\ln n^\text{X}(\va,S_d)}{d^{t_1 }+\big(1+\ln(\va ^{-1})\big)^{t_2}}=0.$$

Clearly, (1,1)-WT is the same as WT, and EC-(1,1)-WT is the same as EC-WT. Obviously, in the  definitions of  SPT, PT, WT and $(t_1,t_2)$-WT, if we replace $\va^{-1}$ by $(1+\ln \va^{-1})$, we get the definitions of EC-SPT, EC-PT, EC-WT and   EC-$(t_1,t_2)$-WT, respectively. We also have
$$\text{SPT} \Longrightarrow \text{PT} \Longrightarrow \text{WT},$$ $$ \text{EC-SPT} \Longrightarrow \text{EC-PT} \Longrightarrow \text{EC-WT},$$
$$\text{EC-SPT} \Longrightarrow \text{SPT},$$ $$\text{EC-PT} \Longrightarrow \text{PT},$$ and $$\text{EC-WT} \Longrightarrow \text{WT}.$$

In the worst case setting the tractability and   EC-tractability of  $L_2$-approximation problems $S_d$
with $G_d=L_2([0,1]^d)$ were investigated in analytic Korobov spaces and weighted Korobov spaces; see \cite{C, DKPW, EP, IKPW, NSW, W, WW, X1}. Additionally, \cite{C,EP,NSW,WW} discussed tractability and EC-tractability  in weighted Korobov spaces.

\begin{lem}\label{le2.1}(\cite{NW1}Theorem 5.2)
Consider the non-zero problem $S=\{S_d\}$ for compact linear problems $S_d$ defined over Hilbert spaces. Then $S$ is PT for NOR iff there exist $q\ge 0$ and $\tau>0$ such that
\begin{equation}\label{e2.2}
C_{\tau,q}:=\sup_{d\in \mathbb N}\big(\sum_{j=1}^\infty(\frac{\lz_{d,j}}{\lz_{d,1}})^\tau\big)^{\frac1 \tau}d^{-q}<\infty.
\end{equation}
Expecially, S is SPT for NOR iff \eqref{e2.2} holds with q=0. The exponent of SPT is
$$p^{\text{str}}=\inf\{2\tau|\tau \ \text{satisfies \eqref{e2.2} with \ } q=0\}.$$
\end{lem}

\subsection{$L_2$-approximation in weighted Korobov spaces and main results}\label{sec2.2}

\

In this subsection we consider $L_2$-approximation  $$\textrm{APP}_d:H_{d,\a,{\bm \gamma}}\to L_2([0,1]^d)$$ with $\textrm{APP}_d(f)=f$ for all $f\in H_{d,\a,{\bm \gamma}}$. Here, the space $H_{d,\a,{\bm \gamma}}$ with weight parameter sequence ${\bm \gamma}=\{\gamma_j\}_{j\in\N}$ and smoothness parameter sequence ${\a}=\{\az_j\}_{j\in\N}$ satisfying \begin{equation}\label{e2.2-2}
1\ge \ga_1\ge \ga_2\ge \cdots\ge 0,
\end{equation}
and \begin{equation}\label{e2.2-3}
 \frac1 2<\az_1\le \az_2\le \cdots.
\end{equation}
is a reproducing kernel Hilbert space. The reproducing kernel function $K_{d,\a,{\g}} :[0,1]^d\times [0,1]^d \to \mathbb{C}$  of the space $H_{d,\a,{\bm \gamma}}$ is given by

$$K_{d,\a,{\g}}(\x,\y) :=\prod_{k=1}^d K_{\az_k,\ga_k}(x_k,y_k),$$
$\x=(x_1,x_2,\cdots,x_d),\,\,\y=(y_1,y_2,\cdots,y_d)\in [0,1]^d$,
where $$K_{\az,\ga}(x,y):=\sum_{k\in \mathbb{Z}}r_{\az,\ga}({k})\exp(2\pi i{k}\cdot(x-y)),\,\;x,\;y\in[0,1]$$ is a universal weighted Korobov kernel function.
Here \begin{equation*}
r _{\alpha,\ga}(k):=\left\{\begin{matrix}
 &1, \,\,\,\,\, \text{ for} \ k=0, \\
 &\frac{\ga}{k^{2\alpha}},\,\,\text{for} \ k\in \mathbb{Z}\setminus \{0\}
\end{matrix}\right.
\end{equation*}
is a positive function and not more than 1 with the smoothness parameter $\alpha>\frac 1 2$. Then we have
$$K_{d,\a,\g}(\x,\y)=\sum_{\k\in \mathbb{Z}^d}r_{d,\a,\g}({\bm k})\exp(2\pi i{\bm k}\cdot(\x-\y)),\;\x, \;\y\in [0, 1]^d,$$ and the corresponding inner product
$$\langle f,g\rangle_{H_{d,\a,\g}}=\sum_{\k\in \mathbb{Z}^d}\frac 1 {r_{d,\a,\g}({\bm k})}\widehat{f}(\k)\overline{\widehat{g}}(\k)$$
and
$$||f||_{H_{d,\a,\g}}=\sqrt{\langle f,f\rangle_{H_{d,\a,\g}}},$$
 where $$r_{d,\a,\g}({\bm k}):=\prod_{j=1}^dr _{\az_j,\ga_j}(k_j),\;\bm k=(k_1,k_2,\dots,k_d)\in \mathbb {Z}^d,$$ $$\x \cdot \y:=\sum_{k=1}^dx_k\cdot y_k,\;\;\x=(x_1,x_2,\dots,x_d),\,\y=(y_1,y_2,\dots,y_d)\in \mathbb{R}^d,$$ and
 $$\widehat{f}(\k)=\int_{[0,1]^d}f(\x)\exp(-2\pi i\k\cdot \x)d\x.$$
Note that the kernel $K_{d,\a,\g}(\x,\y)$ is well defined for all $\frac1 2<\az_1\le \az_2\le \cdots$, since
$$|K_{d,\a,\g}(\x,\y)|\le \sum_{\k\in \mathbb{Z}^d}r_{d,\a,\g}({\bm k})=\prod_{j=1}^d (1+2\ga_j\zeta(2\az_j))<\infty,$$
where $\zeta(\cdot)$ is the Riemann zeta function.
 It is well known from \cite{NSW} that this embedding $\textrm{APP}_d$ is compact with $\frac1 2<\az_1\le \az_2\le \cdots$.

From Subsection \ref{sec2.1} the information complexity of $\text{APP}_d$ depends on the eigenvalues of the operator $W_d=\text{APP}^*_d \text{APP}_d: H_{d,\a,{\bm \gamma}}\to H_{d,\a,{\bm \gamma}}$. Let $(\lz_{d,j}, \eta_{d,j})$ be the eigenpairs of $W_d$,
$$W_d\eta_{d,j}=\lz_{d,j}\eta_{d,j}\,\, \text{for all}\,\, j\in \mathbb{N},$$ where the eigenvalues $\lz_{d,j}$ are ordered,
$$\lz_{d,1}\ge \lz_{d,2}\ge \cdots \ge 0,$$
and the eigenvectors $\eta_{d,j}$ are orthonormal,
$$\langle\eta_{d,i},\eta_{d,j}\rangle_{H_{d,\a,{\bm \gamma}}}=\delta_{i,j}\,\, \text{for all}\,\, i,\, j\in \mathbb{N}.$$
Obviously, we have  $e(0,\text{APP}_d)=1$ (or see \cite{NSW}). Hence the NOR and the ABS for the problem $\text{APP}_d$ coincide in the worst case setting. We set $$n(\va, \text{APP}_d):=n^\text{X}(\va,\text{APP}_d).$$ It is well known that the eigenvalues of the operator $W_d$ are $r_{d,\a,\g}({\bm k})$ with ${\bm k}\in \mathbb{Z}^d$; see, e.g., \cite[p. 215]{NW1}. Hence by \eqref{eq2.2}  we have
\begin{align*}n(\va, \text{APP}_d)&=|\{n\in \mathbb{N}: \lz_{d,n}> \va^{2}\}|=|\{{\bm k}\in \mathbb{Z}^d : r_{d,\a,\g}({\bm k})>\va^{2}\}|\notag\\
&=|\{{\bm k}\in \mathbb{Z}^d :\prod_{j=1}^{d}r _{\alpha_j,\ga_j}(k_j)>\va^{2}\}|.\end{align*}

Tractability such as SPT, PT, WT, and $(t_1,t_2)$-WT for $t_1>1$, and EC-tractability such as EC-WT and EC-$(t_1,1)$-WT for $t_1<1$ of the above problem $\text{APP}=\{\text{APP}_d\}$ with the parameter sequences ${\bm \gamma}=\{\gamma_j\}_{j\in\N}$ and ${\a}=\{\az_j\}_{j\in\N}$ satisfying $$1\ge \ga_1\ge \ga_2\ge \cdots\ge 0$$ and $$\frac 1 2<\az=\az_1= \az_2=\cdots$$ have been solved by \cite{EP, WW} and \cite{C}, respectively. The following conditions have been obtained therein:

$\bullet$ PT  holds iff SPT holds iff $$s_{\g}=\inf\bigg\{\kappa>0: \sum_{j=1}^\infty\ga_j^\kappa<\infty\bigg\},$$
and the exponent  of SPT is $$p^{\text{str}}=2\max\Big(s_{\g},\frac1 {2\az}\Big).$$

$\bullet$ WT holds iff $$\inf_{j\in\N}\ga_j<1.$$

$\bullet$ For $t_1>1$, $(t_1,t_2)$-WT holds for all $1\ge \ga_1\ge \ga_2\ge \cdots\ge 0$.

$\bullet$ EC-WT holds iff $$\lim_{j\to \infty}\ga_j=0.$$

$\bullet$ For $t_1<1$, EC-$(t_1,1)$-WT holds iff $$\lim_{j\to \infty}\frac{\ln j}{\ln(\ga_j^{-1})}=0.$$

We will research  the worst case tractability of the problem APP  with sequences satisfying  \eqref{e2.2-2}  and \eqref{e2.2-3}.

\begin{thm}\label{th1} Consider the above problem $\text{APP}$ with sequences ${\bm \gamma}=\{\gamma_j\}_{j\in\N}$ and ${\a}=\{\az_j\}_{j\in\N}$ satisfying \eqref{e2.2-2} and \eqref{e2.2-3}. Then the problem $\text{APP}$

(1) is SPT or PT  iff
\begin{equation}\label{e1.1}\dz:=\liminf_{j\to\infty}\frac{\ln \ga_j^{-1}}{\ln j}>0.
\end{equation}
The exponent of SPT is $$p^{\text{str}}=2\max\big\{\frac 1 \dz, \frac 1 {2\az_1}\big\}.$$

(2) is $(t_1, t_2)$-WT for all $t_1>1$.

(3) suffers from the curse of  dimensionality with $$\lim_{j\to\infty}\ga_j=1.$$
\end{thm}

\section{The proof}\label{sec3}

\noindent{\it \textbf{Proof of Theorem \ref{th1}.}}

\

(1) For the problem $\textrm{APP}$ we have $\lz_{d,1}=1$. Assume that APP is PT. From Lemma \ref{le2.1} there exist $q\ge 0$ and $\tau>0$ such that
$$C_{\tau,q}:=\sup_{d\in \Bbb N}\big(\sum_{j=1}^\infty\lz_{d,j}^\tau\big)^{\frac1 \tau}d^{-q}<\infty.$$
Since \begin{equation}\label{e3.1-1}\sum_{j=1}^\infty\lz_{d,j}^\tau=\prod_{j=1}^d\big(1+2\ga_j^\tau\zeta(2\az_j\tau)\big),\end{equation}
we have
\begin{align*}C_{\tau,q}&\ge \big(\sum_{j=1}^\infty\lz_{d,j}^\tau\big)^{\frac1 \tau}d^{-q}\\
&=\big[\prod_{j=1}^d\big(1+2\ga_j^\tau\zeta(2\az_j\tau)\big)\big]^{\frac1 \tau}d^{-q}\\
&\ge [\prod_{j=1}^d\big(1+\ga_j^\tau\big)^{\frac1 \tau}]d^{-q}\\
&\ge(1+\ga_d^\tau)^{\frac d \tau}d^{-q}.\end{align*}
It follows that
$$\ln C_{\tau,q}+q\ln d\ge \frac d \tau\ln(1+\ga_d^\tau)\ge \frac d \tau \cdot\frac {\ga_d^\tau}{2},$$
where we used $\ln(1+x)\ge \frac x 2$ for all $x\in[0,1]$. We further get
$$\ln(\ln C_{\tau,q}+q\ln d)\ge \ln d -\tau\ln \ga_d^{-1}-\ln (2\tau),$$
i.e., $$\frac{\ln \ga_d^{-1}}{\ln d}\ge \frac{\ln d-\ln(\ln C_{\tau,q}+q\ln d)-\ln 2\tau}{\tau\cdot\ln d}.$$
Hence we obtain \begin{equation}\label{e3.1}\dz={\lim\inf}_{d\to\infty}\frac{\ln \ga_d^{-1}}{\ln d}\ge \frac 1 \tau>0.\end{equation}
Note that if APP is SPT, then it is PT. It implies that if APP is SPT, then \eqref{e3.1} holds and the exponent $$p^{\text{str}}\ge 2\max\{\frac 1 \dz, \frac 1 {2\az_1}\}.$$

On the other hand, assume that \eqref{e1.1} holds. For an arbitrary $\va\in(0,\frac \dz 2)$, there exists an integer $N>0$ such that for all $j\ge N$
we have $$\frac{\ln\ga_j^{-1}}{\ln j}\ge \dz-\va.$$ It means that for all $j\ge N$ $$\ga_j\le j^{-(\dz-\va)}.$$
Choosing $\tau=\frac 1 {\dz-2\va}$ and noting that  $\frac{\dz-\va}{\dz-2\va}>1$, we have
$$\sum_{j=N}^\infty\ga_j^\tau\le\sum_{j=N}^\infty j^{-(\dz-\vz)\tau}=\sum_{j=N}^\infty j^{-\frac{\dz-\vz}{\dz-2\vz}}<\infty,$$
which yields that \begin{equation}\label{e3.2}
\sum_{j=1}^\infty\ga_j^\tau<\infty.
\end{equation}
From \eqref{e3.2} we get
\begin{align*}
\big(\sum_{j=1}^\infty\lz_{d,j}^\tau\big)^{\frac 1 \tau}d^{-q}&=\big[\prod_{j=1}^d\big(1+2\ga_j^\tau\zeta(2\az_j\tau)\big)\big]^{\frac1 \tau}d^{-q}\\
&\le \big[\prod_{j=1}^d\big(1+2\ga_j^\tau\zeta(2\az_1\tau)\big)\big]^{\frac1 \tau}d^{-q}\\
&=d^{-q}\cdot\exp\Big\{\ln\big[\prod_{j=1}^d\big(1+2\ga_j^\tau\zeta(2\az_1\tau)\big)^{\frac1 \tau}\big]\Big\}\\
&=d^{-q}\cdot\exp\Big\{\frac1 \tau\sum_{j=1}^d\ln\big[\big(1+2\ga_j^\tau\zeta(2\az_1\tau)\big)\big]\Big\}\\
&\le d^{-q}\cdot\exp\Big\{\frac1 \tau\sum_{j=1}^d2\ga_j^\tau\zeta(2\az_1\tau)\Big\}\\
&= d^{-q}\cdot\exp\Big\{\frac {2\zeta(2\az_1\tau)} {\tau}\cdot\sum_{j=1}^d\ga_j^\tau\Big\}\\
&\le d^{-q}\cdot\exp\Big\{\frac {2\zeta(2\az_1\tau)} {\tau}\cdot\sum_{j=1}^\infty\ga_j^\tau\Big\}\\
&<\infty
\end{align*}
for any $q\ge0$ and $\tau>\frac 1 {2\az_1}$. We further get
\begin{align*}C_{\tau,q}&=\sup_{d\in \mathbb N}\big(\sum_{j=1}^\infty\lz_{d,j}^\tau\big)^{\frac1 \tau}d^{-q}<\infty\end{align*}
for any $q\ge0$ and $\tau>\frac 1 {2\az_1}$.
It follows from Lemma \ref{le2.1} that APP is SPT or PT and the exponent $p^{\text{str}}\le 2\tau$. Setting $\va\to 0$, we obtain
$$p^{\text{str}}\le 2\tau \le2\max\big\{\frac 1 \dz, \frac 1 {2\az_1}\big\}.$$
Hence the exponent of SPT is $p^{\text{str}}=2\max\big\{\frac 1 \dz, \frac 1 {2\az_1}\big\}.$

\

(2) Due to $$n\lz_{d,n}^\tau\le\sum_{j=1}^n\lz_{d,j}^\tau\le\sum_{j=1}^\infty\lz_{d,j}^\tau,$$
we have $$\lz_{d,n}\le\frac{(\sum_{j=1}^\infty\lz_{d,j}^\tau)^{\frac1 \tau}}{n^{\frac 1 \tau}}.$$
Noting that $\lz_{d,n}\le \va^2$ holds for
$$n=\lceil\sum_{j=1}^\infty\lz_{d,j}^\tau\va^{-2\tau}\rceil,$$ we get
\begin{align*}n(\va, \text{APP}_d)&=\min\{n| \lz_{d,n+1}\le \va^{2}\}\\
&\le \lceil\sum_{j=1}^\infty\lz_{d,j}^\tau\va^{-2\tau}\rceil\\
&\le1+\va^{-2\tau}\sum_{j=1}^\infty\lz_{d,j}^\tau\\
&=1+\va^{-2\tau}\prod_{j=1}^d\big(1+2\ga_j^\tau\zeta(2\az_j\tau)\big)\\
&\le 2\va^{-2\tau}\prod_{j=1}^d\big(1+2\ga_j^\tau\zeta(2\az_j\tau)\big),\end{align*}
where we used \eqref{e3.1-1} in the last equality.
It follows that \begin{align*}\frac{\ln n(\va, \text{APP}_d)}{d^{{t_1}}+\va^{-t_2}}&\le\frac{\ln\Big(2\va^{-2\tau}\prod_{j=1}^d\big(1+2\ga_j^\tau\zeta(2\az_j\tau)\big)\Big)}{d^{{t_1}}+\va^{-t_2}}\\
&=\frac{\ln2 +2\tau\ln(\va^{-1})+\sum_{j=1}^d\ln\big(1+2\ga_j^\tau\zeta(2\az_j\tau)\big)}{d^{{t_1}}+\va^{-t_2}}\\
&\le\frac{\ln2 +2\tau\ln(\va^{-1})+\sum_{j=1}^d2\ga_j^\tau\zeta(2\az_j\tau)}{d^{{t_1}}+\va^{-t_2}}\\
&\le\frac{\ln2 +2\tau\ln(\va^{-1})+2\zeta(2\az_1\tau)\sum_{j=1}^d\ga_j^\tau}{d^{{t_1}}+\va^{-t_2}}\\
&\le\frac{\ln2 +2\tau\ln(\va^{-1})+2d\zeta(2\az_1\tau)}{d^{{t_1}}+\va^{-t_2}}\\
&\le\frac{\ln2 +2\tau\ln(\va^{-1})}{\va^{-t_2}}+\frac{2d\zeta(2\az_1\tau)}{d^{{t_1}}}.\end{align*}
We obtain for all $t_1>1$ and $t_2>0$, $$\lim_{\varepsilon ^{-1}+d\rightarrow \infty}\frac{\ln n(\va, \text{APP}_d)}{d^{{t_1}}+\va^{-t_2}}=0,$$
which means APP is $(t_1,t_2)$-WT for all $t_1>1$ and $t_2>0$.

\

(3) Since $1\ge \ga_1\ge \ga_2\ge \cdots\ge 0$, we have for all $j\in {\mathbb N}$ $$1=\lim_{j\to\infty}\ga_j = \inf_{j\in {\mathbb N}}\ga_j\le\ga_j\le1,$$
which implies that $\ga_j\equiv1$ for all $j\in {\mathbb N}$. It follows that $$1=r_{d,\a,\g}({\bm k})>\va^{2}$$ for all ${\bm k}\in\{0,1,-1\}^d$. Then we have $$n(\va, \text{APP}_d)=|\{{\bm k}\in \mathbb{Z}^d : r_{d,\a,\g}({\bm k})>\va^{2}\}|\ge 3^d.$$ Hence APP suffers from the curse of  dimensionality.
$\hfill\Box$

\

In this paper we consider the SPT, PT and $(t_1, t_2)$-WT for all $t_1<1$ and $t_2>0$ for worst case $L_2$-approximation in weighted Korobov spaces $H_{d,\a,{\bm \gamma}}$ with parameters $1\ge \ga_1\ge \ga_2\ge \cdots\ge 0$ and $\frac1 2<\az_1\le \az_2\le \cdots$. We get the matching necessary and sufficient condition on SPT or PT. In particular, it is $(t_1, t_2)$-WT for all $t_1>1$ and $t_2>0$ and suffers from the curse of  dimensionality with $\lim_{j\to\infty}\ga_j=1.$

\

\section*{Acknowledgments}
 This work was supported by the National Natural Science Foundation of China (Project no. 12001342), Scientific and Technological Innovation Project of Colleges and Universities in Shanxi Province (Project no. 2022L438), Basic Youth Research Found Project of Shanxi Datong University (Project no. 2022Q10), Doctoral Foundation Project of Shanxi Datong University (Project no. 2019-B-10) and Doctoral Foundation Project of Shanxi Datong University (Project no. 2021-B-17).

\end{document}